\documentclass{aa}
\usepackage{txfonts}
\usepackage{graphicx}
\usepackage[latin1]{inputenc}
\begin{document}
\def\Hp{H\,{\sc{ii}}}
\def\SII{[S\,{\sc{ii}}]}
\def\Ha{H$\alpha$}
\def\FeII{[Fe\,{\sc{ii}}]}
\def\HI{H\,{\sc{i}}}
\def\fs{\hbox{$.\!\!^{\rm s}$}}
\def\fdg{\hbox{$.\!\!^\circ$}}
\def\farcm{\hbox{$.\mkern-4mu^\prime$}}
\def\farcs{\hbox{$.\!\!^{\prime\prime}$}}
\def\arcmin{\hbox{$^\prime$}}
\def\arcsec{\hbox{$^{\prime\prime}$}}
\def\sun{\hbox{$\odot$}}
\def\degr{\hbox{$^\circ$}}
\def\h{\hbox{$^{\reset@font\r@mn{h}}$}}
\def\m{\hbox{$^{\reset@font\r@mn{m}}$}}
\def\s{\hbox{$^{\reset@font\r@mn{s}}$}}

\def\msol{\hbox{\kern 0.20em $M_\odot$}}
\def\lsol{\hbox{\kern 0.20em $L_\odot$}}
\def\smu{\hbox{\kern 0.20em s$^{-1}$}}
\def\kms{\hbox{\kern 0.20em km\kern 0.20em s$^{-1}$}}
\def\cmmt{\hbox{\kern 0.20em cm$^{-3}$}}
\def\cmmd{\hbox{\kern 0.20em cm$^{-2}$}}
\def\K{\hbox{\kern 0.20em K}}
\def\pc{\hbox{\kern 0.20em pc}}
\def\pcpd{\hbox{\kern 0.20em pc$^{2}$}}
\def\pcmu{\hbox{\kern 0.20em pc$^{-1}$}}
\def\twco{\hbox{${}^{12}$CO}}
\def\twcotwo{\hbox{${}^{12}$CO(2-1)}}
\def\thco{\hbox{${}^{13}$CO}}
\def\thcotwo{\hbox{${}^{13}$CO(2-1)}}
\def\thcoone{\hbox{${}^{13}$CO(1-0)}}
\def\ceio{\hbox{C${}^{18}$O}}
\def\ceiotwo{\hbox{C${}^{18}$O(2-1)}}
\def\ceioone{\hbox{C${}^{18}$O(1-0)}}
\def\cs{\hbox{CS}}
\def\csthree{\hbox{CS(3-2)}}
\def\cstwo{\hbox{CS(2-1)}}
\def\csfive{\hbox{CS(5-4)}}
\def\cts{\hbox{C${}^{34}$S}}
\def\ctsthree{\hbox{C${}^{34}$S(3-2)}}
\def\ctstwo{\hbox{C${}^{34}$S(2-1)}}
\def\htwo{\hbox{H${}_2$}}
\def\h13cop{\hbox{H$^{13}$CO$^{+}$}}
\def\halpha{\hbox{H$\alpha$ }}
\def\hcop{\hbox{HCO$^{+}$}}
\newcommand{\jonetozero}{\hbox{$J=1\rightarrow 0$}}
\newcommand{\jtwotoone}{\hbox{$J=2\rightarrow 1$}}
\newcommand{\jthreetotwo}{\hbox{$J=3\rightarrow 2$}}
\newcommand{\jfourtothree}{\hbox{$J=4\rightarrow 3$}}
\newcommand{\jfivetofour}{\hbox{$J=5\rightarrow 4$}}
\title{Triggered massive-star formation on the borders of 
       Galactic H\,{\Large II} regions }
\subtitle{IV. Star formation at the periphery of Sh2-212 \thanks{Based 
on observations obtained at the IRAM, Spain, at the CFHT, Hawaii, 
at the VLA, USA, and at the Observatoire de Haute-Provence, France.} 
{\mbox{\,\,}}\thanks{Table~3 is available in electronic form at the CDS 
via anonymous ftp to cdsarc.u-strasbg.fr (130.79.128.5) or via 
http://cdsweb.u-strasbg.fr/cgi-bin/qcat?J/A+A/}}
\author{L.~Deharveng\inst{1}
          \and
	B.~Lefloch\inst{2}
         \and
        S.~Kurtz\inst{3}
           \and
	D.~Nadeau\inst{4}
	    \and
        M.~Pomar\`es\inst{1}
           \and
        J.~Caplan\inst{1}
	   \and
	A.~Zavagno\inst{1}
        }
        
\offprints{L.~Deharveng}

\institute{Laboratoire d'Astrophysique de Marseille, 2 place Le Verrier, 
13248 Marseille Cedex~4, France, lise.deharveng@oamp.fr
      \and
       Laboratoire d'Astrophysique de l'Observatoire de Grenoble, 414 rue de la Piscine, 
       BP 53, 38041 Grenoble Cedex 9, France
      \and
       Centro de Radioastronom\'{i}a y Astrof\'{i}sica, UNAM, Apartado Postal 3-72, 
       58089, Morelia, Michoac\'an, M\'exico
      \and
       Observatoire du Mont M\'egantic et D\'epartement de Physique, Universit\'e de Montr\'eal, 
       CP~6128, Succ. Centre-Ville, Montr\'eal, QC, Canada, H3C3J7
	}
	
\date{Received; accepted }

\abstract{}
{We wish to establish whether sequential star formation is taking place at 
the periphery of  the Galactic \Hp\ region Sh2-212.} 
{We present CO millimetre observations of this region obtained at the IRAM 
30-m telescope to investigate the distribution of associated molecular material. 
We also use deep $JHK$ observations obtained at the CFHT to 
study the stellar content of the region, and radio observations obtained 
at the VLA to look for the presence of an ultra-compact (UC) \Hp\ region 
and for maser emission.}
{In the optical, Sh2-212 is spherically symmetric around its central exciting 
cluster. This \Hp\ region is located along a molecular filament. A thin, 
well-defined half ring of molecular material surrounds the brightest part of 
the \Hp\ region at the rear and is fragmented. The most massive 
fragment ($\sim$200\,\msol) contains a massive young stellar object displaying 
a near-IR excess; its spectral energy distribution indicates a high-mass 
($\sim$14\,\msol), high-temperature ($\sim$30000\,K), and high-luminosity 
($\sim$17000\,\lsol) source. This object ionizes a UC \Hp\ region.}
{Sh2-212 is a good example  of massive-star formation triggered via the 
collect and collapse process. The massive YSO observed at its periphery is a 
good candidate for a massive star formed in isolation.}

\keywords{Stars: formation -- Stars: early-type -- ISM: \Hp\ regions --
   ISM: individual: Sh2-212}

 \titlerunning{Star formation near Sh2-212}
 \authorrunning{L.~Deharveng, B.~Lefloch et al.}
 
\maketitle

\section{Introduction}

Various mechanisms may trigger star formation on the borders of \Hp\ regions 
(see the review by Elmegreen~\cite{elm98}). All rely on the high-pressure
exerted by the warm ionized gas on the surrounding cold neutral material. 
These mechanisms differ in their assumptions concerning the nature of 
the surrounding medium (homogeneous or not) and the part played by turbulence. 
One of these mechanisms, the collect and collapse process, first proposed by 
Elmegreen \& Lada~(\cite{elm77}), is particularly interesting as it allows 
the formation of massive fragments (hence subsequently of massive objects, 
stars or clusters), out of an initially uniform medium. In this process 
a layer of neutral material is collected between the ionization front (IF) and 
the associated shock front (SF) during the supersonic expansion of an \Hp\ region. 
With time this layer may become massive and gravitationally instable, leading 
to the formation of dense massive cores (Whitworth et al.~\cite{whi94}, 
Hosokawa \& Inutsuka~\cite{hos06}). 

We have previously proposed seventeen Galactic \Hp\ regions as 
candidates for the collect and collapse process of massive-star formation 
(Deharveng et al.~\cite{deh05}). Among these is Sh2-212, the subject of the 
present paper, described in Sect.~2. Our main criterion for the 
choice of Sh2-212 was the presence of a bright MSX point source 
(Price et al.~\cite{pri01}) at its periphery, beyond the 
ionization front, coincident with a red object in the 2MASS survey  
(Skrutskie et al.~\cite{skr06}) and 
with a small optical reflection nebula. However nothing was yet known about 
Sh2-212's molecular environment. We now present new high-resolution molecular 
observations to investigate the distribution of molecular material. 
Do we observe a layer of dense neutral material surrounding the ionized 
gas -- a signature of the collect and collapse process of star formation? 
These 
observations are described in Sect.~4.  We also present new $JHK$ observations 
to determine the stellar content of this region. Are young stellar 
objects (YSOs) present on the border of Sh2-212? What is the nature of the 
MSX point source 
observed near the IF? This is discussed in Sect.~3. We also 
present new radio observations aimed at detecting possible UC \Hp\ regions at 
the periphery of Sh2-212, and at detecting maser 
emission indicative of recent star formation. These 
observations are described in Sect.~5.

The results are discussed in Sect.~6, where we present our view of the 
morphology of the whole complex, and argue in favour of the 
collect and collapse process of massive-star formation. \\

\section{Description of the region}

Sh2-212 (Sharpless~\cite{sha59}) is a bright optically-visible \Hp\ region in 
the outer Galaxy ({\it l}=155\fdg36, {\it b}=2\fdg61). It lies high 
above the Galactic plane ($\sim300$~pc assuming a distance of 6.5~kpc, Sect.~2.1 ), 
and far from the Galactic centre (14.7~kpc). Its diameter is 
$\sim$5\arcmin\ (9.5~pc). It is a high-excitation \Hp\ region, 
ionized by a cluster containing an O5.5neb (Moffat et 
al.~\cite{mof79}) or an O6I star (Chini \& Wink~\cite{chi84}). 

Fig.~1 presents a colour image of Sh2-212 in the optical, a 
composite of two frames obtained at the 120-cm telescope of the Observatoire 
de Haute-Provence. Pink corresponds to the \Ha\ emission at 6563~\AA\ (exposure 
time 1~hour) and turquoise to the \SII\ emission at 6717~\AA\ and 6731~\AA\ 
(exposure time 2$\times$1~hour). \SII\ is enhanced near the 
ionization front, and thus is a good tracer of the limits of the ionized 
region and of the morphology of the ionization front. 
Sh2-212 appears as a circular \Hp\ region around its exciting cluster. 
Numerous substructures are present, indicating that Sh2-212 is presently 
evolving in a non-homogeneous medium. A bright rim is conspicuous at the 
north-western border of Sh2-212. A small reflection nebula (indicated by an 
arrow in Fig.~1) is present beyond this ionization front.

Because Sh2-212 is both optically bright and situated far (14.7~kpc) from the 
Galactic centre, it has been included in numerous studies of abundance 
determinations in the Galaxy. For this purpose absolute integrated line fluxes in a 
number of nebular emission lines were measured through a circular diaphragm 
by Caplan et al.~(\cite{cap00}). These measurements confirm the  
high-excitation of Sh2-212; they indicate an electron temperature of 9700~K  
and an electron density of 130~cm$^{-3}$ (Deharveng et al.~\cite{deh00}).

The coordinates of the objects discussed in the text are given in Table~1.

%
\begin{figure}[htp]
\includegraphics[width=90mm]{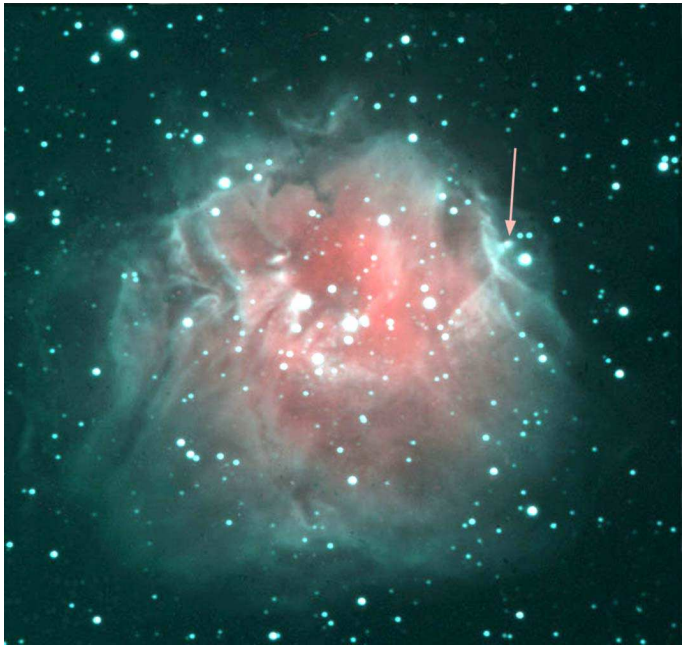}
\caption{Composite colour image of Sh2-212 in the optical. 
North is up and east is left. The size of the field is 
$7\farcm0$ (E-W) $\times$ $6\farcm6$ (N-S).
Pink corresponds to the H$\alpha$ 6563\AA\ emission, and turquoise to the 
\SII\  6717\AA\ + 6731\AA\ emission, enhanced near the ionization front. The 
arrow points to the reflection nebulosity, associated with star no.~228,  
discussed in the text.} 
\end{figure}

\begin{table}[htp]
\caption{Coordinates of the objects discussed in the text}
\begin{tabular}{lllllll}
 \hline\hline
 Object & \multicolumn{3}{c}{RA~(2000)} & \multicolumn{3}{c}{Dec~(2000)}\\
 & h & m & s & $\degr$ & $\arcmin$ & $\arcsec$ \\
  \hline
 Main exciting star M2 & 4 & 40 & 37.44 & +50 & 27 & 40.5 \\
 MSX G155.3319+02.5989 & 4 & 40 & 27.2 & +50 & 28 & 29 \\
 IRAS 04366+5022       & 4 & 40 & 26.1 & +50 & 28 & 24 \\
 UC \Hp\ region        & 4 & 40 & 27.2 & +50 & 28 & 29 \\     
 Massive YSO no.~228     & 4 & 40 & 27.24 & +50 & 28 & 29.5 \\ 
 \hline \\
\end{tabular}

\end{table}

Sh2-212 is a thermal radio-continuum source, with a flux density of 1.58~Jy 
at 1.46~GHz (Fich~\cite{fic93}, and references therein). The angular 
resolution of Fich's observations, 40\arcsec, was insufficient for the  
detection of a possible UC \Hp\ region on the border of Sh2-212. 
Higher angular resolution radio observations will be presented 
and discussed in Sect.~5.

Table~2 lists the velocities, obtained by various authors, of the ionized 
gas and the associated molecular material. As a whole, the ionized gas flows 
away from the molecular cloud, with a radial velocity of the order of 
5~km~s$^{-1}$. This may be indicative of a ``champagne flow'' 
(Tenorio-Tagle~\cite{ten79}). This point will be developed in Sect.~6.

\begin{table}[htp]
\caption{Velocity measurements}
\begin{tabular}{lllllll}
 \hline\hline
 Line & V$_{\rm LSR}$ (km s$^{-1}$) & Reference \\
 \hline
 H$\alpha$ & $-$39.5   & Pi{\c{s}}mi{\c{s}} et al.~(\cite{pis91}) \\
 H$\alpha$ & $-$43.9$\pm$0.1 & Fich et al.~(\cite{fic90}) \\
 H109$\alpha$ & $-$40.1$\pm$1.0 & Lockman~(\cite{loc89}) \\
 CO & $-$35.3$\pm$0.3 & Blitz et al.~(\cite{bli82}) \\
 CO & $-$34.3 & Shepherd \& Churchwell~(\cite{she96})\\
 \hline \\
\end{tabular}

\end{table}

Sh2-212 was proposed by Deharveng et al.~(\cite{deh05}) as a candidate for the 
collect and collapse process of massive-star formation, on the basis of  
i)~the presence of a ring of MSX emission at 8.3~$\mu$m 
(mainly PAH emission) surrounding the 
brightest part of the ionized region; this indicates 
the presence of dense neutral material and dust around the ionized gas; and ii) the 
presence of a luminous MSX point source in the direction of this dust ring 
(indicated by an arrow in Fig.~2). This MSX point source lies in the 
direction of the reflection nebulosity. Fig.~2 shows that the bright ring of 
MSX emission at 8.3~$\mu$m surrounds only the bright northern part of the ionized 
region, and not the whole region. However, fainter brightness emission is 
observed around the southern part of the \Hp\ region. This point will be 
discussed in Sect.~6.

%
\begin{figure}
\includegraphics[width=90mm]{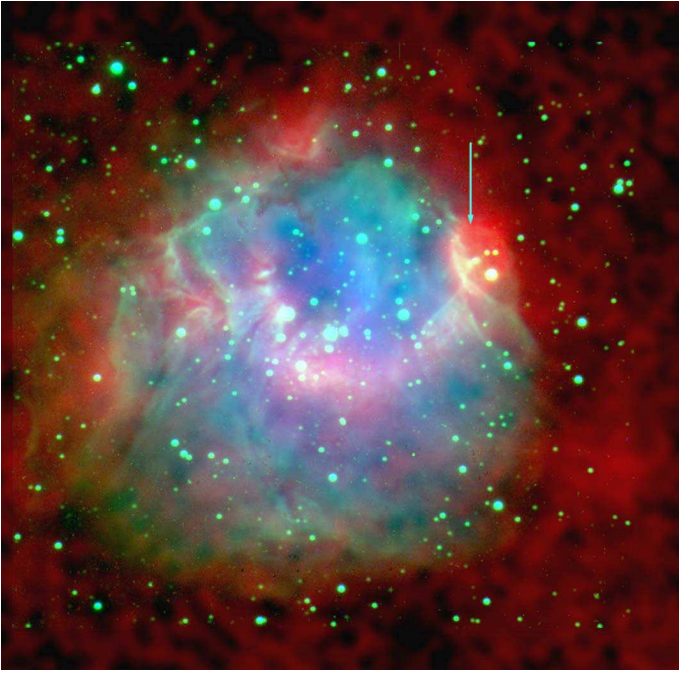}
\caption{Composite colour image of Sh2-212 in the optical and the mid-IR. 
Red corresponds to the MSX emission at 8.3~$\mu$m, and turquoise to the optical 
emission of the ionized gas. The arrow points to the MSX point source 
associated with star no.~228 and the UC \Hp\ region discussed in the text.} 
\end{figure}

\subsection{The distance of Sh2-212}

A kinematical distance $D$ can be estimated from the velocity of the molecular 
gas, $V_{\rm LSR}(\rm CO)=-35 \pm 1$~km~s$^{-1}$ (Table~2), and from the Galactic 
rotation curve of Brand \& Blitz~(\cite{bra93}); we obtain $D=6.1 \pm0.4$~kpc.

A photometric distance can be estimated for M2, the main exciting star of Sh2-212. 
We have used the spectral type and the $UBV$ magnitudes of Moffat et 
al.~(\cite{mof79}; O5.5V, $U$=11.90, $B$=12.34, $V$=11.77), our 
$JHK$ magnitudes (Sect.~4; $J$=10.38, $H$=10.17, $K$=10.07), the synthetic 
photometry of O stars by Martins \& Plez (\cite{mar06}), 
and the interstellar extinction law of 
Rieke \& Lebofsky~(\cite{rie85}). The best fit is obtained for a distance of 
6.5~kpc and a visual extinction $A_V$ of 2.85~mag.

In the following we adopt this distance of 6.5~kpc, which is consistent  
with the kinematic one.

\section{Near IR observations}
\subsection{Observations}

Sh2-212 was observed with the CFHT-IR camera on the night of 2002 October 20. 
Frames were obtained in the $JHK$ broad-band filters. The 
detector was a Rockwell array of 1024$\times$1024 pixels, with a pixel size of  
$0\farcs211$. For each band a mosaic of nine positions was obtained, each position being 
observed ten times with a short exposure time. This results in a 
total field of view of $4\farcm4$ (E-W) $\times$ $5\farcm2$ (N-S), and  
total integration times of 270s, 270s, and 450s in the $J$, $H$, 
and $K$ bands respectively. 

The $J$, $H$, and $K$ images were reduced using the DAOPHOT stellar 
photometry package (Stetson~\cite{ste87}), with PSF fitting. The results 
were calibrated using the 2MASS Point Source Catalog 
(Skrutskie et al. \cite{skr06}), with 65 common stars. 
After a best-fit transformation, an rms dispersion of 0.10~mag is present 
between our photometry and 2MASS, in each band and each colour. A total 
of 891 sources were measured in the three $JHK$ bands, and 36 more
were measured in only one or two bands. The detection 
limit is $\sim$17.5~mag in $J$ and $K$, and $\sim$18~mag in $H$. 
The seeing was $0\farcs9$.

Fig.~3 presents a composite $JHK$ colour image of Sh2-212. Our observations do 
not cover the whole \Hp\ region (our field is centred on the massive YSO,  
star no.~228, discussed below). We have supplemented the coverage of the 
Sh2-212 field, when necessary, using the 2MASS survey. Table~3, giving 
the coordinates and the magnitudes of the stars in the $JHK$ bands 
(CFHT observations), is available in electronic form at the CDS.

%
\begin{figure}
\includegraphics[width=90mm]{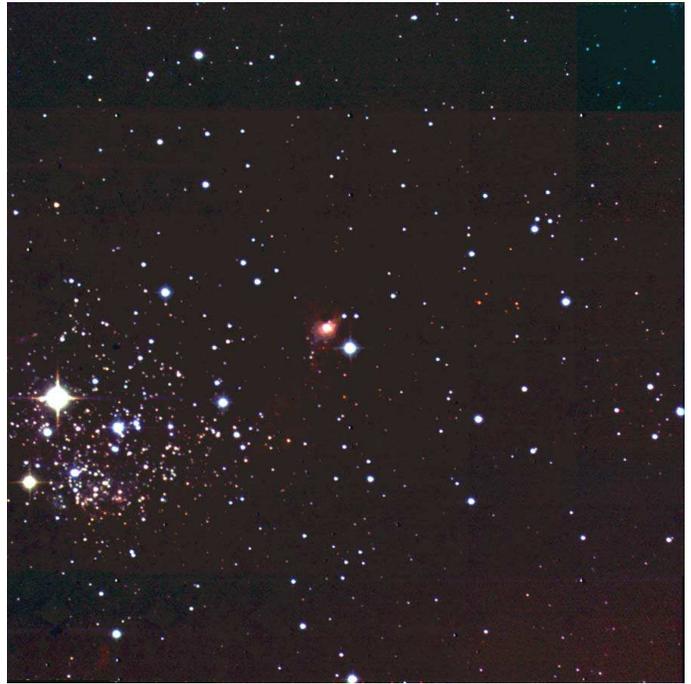}
\caption{Sh2-212. Composite colour image of Sh2-212 in the near-IR ($J$ is 
blue, $H$ is green, $K$ is red). The colours of the stars are mainly 
determined by extinction. 
North is up and east is left. The size of the field is 
$4\farcm4$ (E-W) $\times$ $5\farcm2$ (N-S). The red nebulosity at the centre   
of the field corresponds to the optical reflection nebula associated with 
the MSX point source and star no.~228.}
\end{figure}

\begin{table*}[htp]
\caption{ Coordinates and $JHK$ photometry of all the stars in a 
$4\farcm4 \times 5\farcm2$ field centred on star no.~228. This table 
is available in electronic form at the CDS 
via anonymous ftp to cdsarc.u-strasbg.fr (130.79.128.5) or via 
http://cdsweb.u-strasbg.fr/cgi-bin/qcat?J/A+A/.}
\end{table*}

\subsection{The stellar population associated with Sh2-212}

Fig.~4 presents the $K$ versus $J-K$ magnitude-colour diagram  
of the sources detected in the three bands. The main sequence 
is drawn for a distance of 6.5~kpc, using the absolute calibration 
and colours of Martins et al.~(\cite{mar05}) and Martins \& 
Plez~(\cite{mar06})  
for O3 to O9.5 stars; for later spectral types the absolute calibration 
is that of Schmidt-Kaler~(\cite{sch82}) and the colours are from 
Tokunaga~(\cite{tok00}). Note that there is some overlap between the
 magnitudes of O9.5 and B0 stars. The reddening lines correspond to a visual 
extinction of 30~mag. The interstellar extinction law is from Indebetouw et 
al.~(\cite{ind05}). 

The $J-H$ versus $H-K$ colour-colour diagram is presented in Fig.~5. The 
reddening lines are drawn for a visual extinction of 20~mag. They bracket the 
region occupied by reddened main-sequence stars. Stars near or above the 
upper reddening line may be evolved stars (giants). Stars below the bottom 
reddening line have a near-IR excess; they are probably young stellar objects 
associated with large amounts of dust (in an envelope or a disk), such as 
T~Tauri stars, Herbig Ae/Be stars and more massive YSOs (Lada \& Adams~\cite{lad92}).

We have labelled a few objects in Fig.~6:

$\bullet$ the stars observed in $UBV$ by Moffat et al.~(\cite{mof79}), each 
marked with the letter M followed by the number given by these authors.

$\bullet$ the stellar object observed in the direction of the MSX and IRAS 
point sources. The large symbols in Figs 4 and 5 correspond to the magnitudes 
and colours of the whole object (star no.~228 plus associated nebulosity) 
integrated in a diaphragm of radius 6$\farcs$3. Note that this object 
presents a near-IR colour excess. 
 
$\bullet$ a few other stars, saturated on our frames, are identified by 
asterisks; their magnitudes are from the 2MASS catalogue.

Fig.~4 shows that the whole region 
is affected by a visual extinction of about 3~mag, most probably of 
interstellar origin. Moffat et al.'s stars are, at optical wavelengths, the 
brightest components in the observed field;  most of these stars 
follow the main sequence reddened by $\sim$3~mag. In particular,  this is the 
case of M2, an O5.5 star according to Moffat et al., and the main exciting 
star of Sh2-212. Our $JHK$ photometry confirms this conclusion.  
However, Figs~4 and 5 show that a few of Moffat et al.'s stars are evolved; 
this is the case for stars M6, M10, and possibly M8 and M9. Star A, very 
bright in the near IR but not in the optical (not measured by Moffat et al.),  
is also an evolved star. (Fig.~4 shows that it is too luminous 
to be a main-sequence star associated with Sh2-212, and Fig.~5 shows it to be 
a giant star.) It is not possible to know whether  these stars 
belong to the exciting cluster or if they are unrelated foreground stars. 
Our near-IR images also show that stars M1 and M3 are double stars. 

Red stars are observed all over the field of Fig.~3. Their density is 
especially high in the direction or in the vicinity of the molecular 
condensation C2 (Sect. 4.2). Many of these objects present a near-IR 
excess, indicating that they are YSOs.
 
The extinction affecting the central cluster of Sh2-212 
(about 3 mag) is very low for the 
distance of the region and is thus probably mainly of interstellar origin. 
Thus very little {\it local} dust is present in front of the optical nebula 
and its exciting cluster. 

%
\begin{figure}[htp]
\includegraphics[width=90mm]{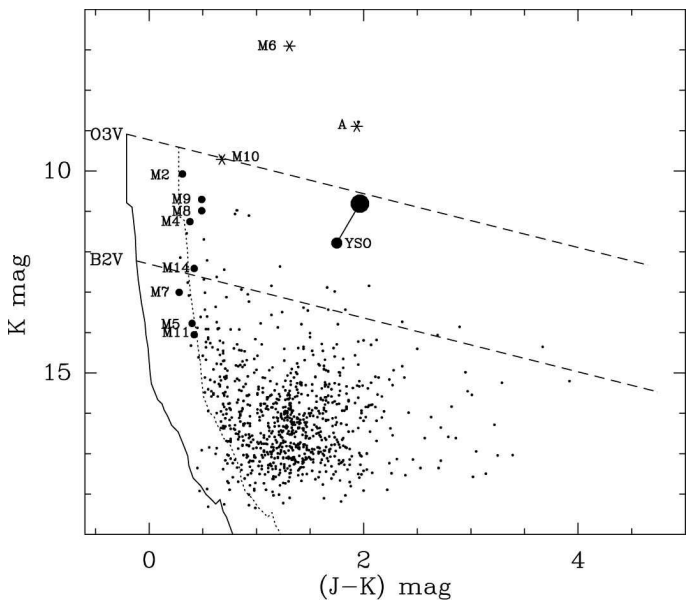}
\caption{$K$ versus $J-K$ diagram. The main sequence is drawn for a 
visual extinction of zero (full line) and 3~mag (dotted line). The reddening 
lines, corresponding to a visual extinction of 30~mag, are issued from O3V 
and B2V stars. A few stars are identified, according to Fig.~6. Moffat et 
al.'s (\cite{mof79}) stars are identified by their number according to these 
authors. The asterisks are for 2MASS measurements. The connected full circles 
are for the star no.~228 (the YSO) and its associated 
nebulosity.}
\end{figure}

%
\begin{figure}[htp]
\includegraphics[width=90mm]{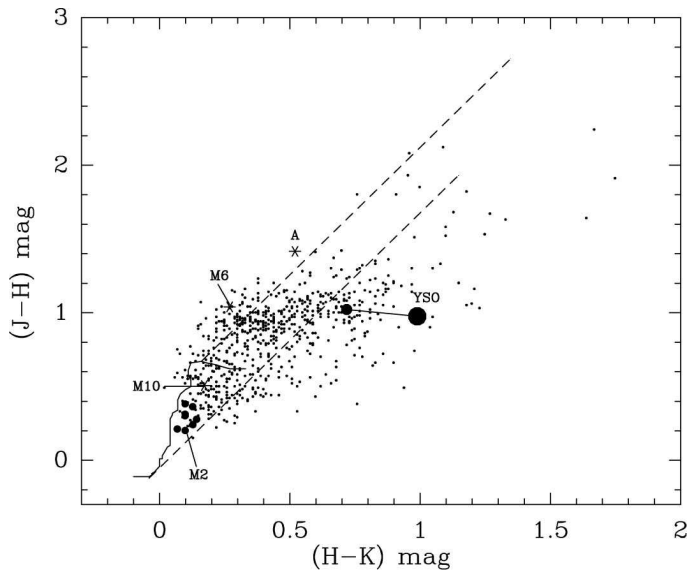}
\caption{$J-H$ versus $H-K$ diagram for stars with $K\leq17$~mag. The main 
sequence (full line) is drawn for a visual extinction of zero. The reddening 
lines, starting at the positions of B0V and M0V stars, have lengths 
corresponding to a visual extinction of 20~mag. The symbols are the same as 
in Fig.~4.}
\end{figure}

%
\begin{figure}[htp]
\includegraphics[width=90mm]{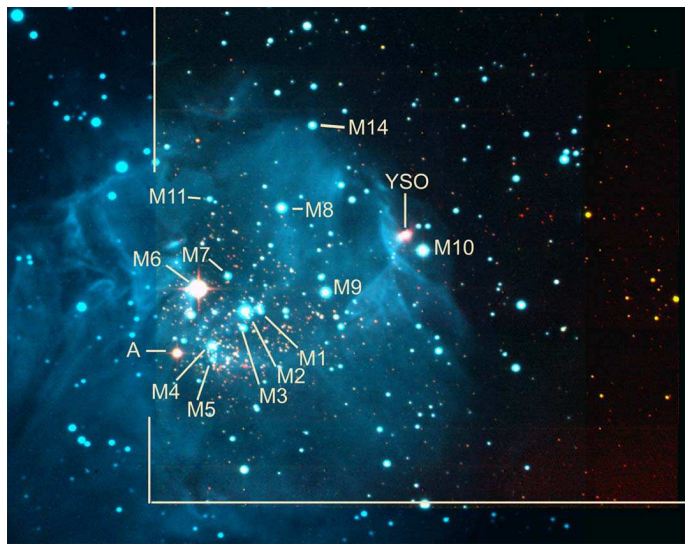}
\caption{Identification of a few objects discussed in the text. The underlying
 image is a colour composite of the \SII\ frame (blue) and of the $K$ frame 
(orange). The lines indicate the limits of our $JHK$ frames.}
\end{figure}

\section{Molecular observations}
\subsection{Observations}

In March 2006 we observed the emission of the molecular gas associated 
with Sh2-212, in the \twco\ and \thco\ \jtwotoone\  lines  using the IRAM 
30-m telescope (Pico Veleta, Spain). We mapped an area of 
$16\arcmin \times 16\arcmin$ with the HERA heterodyne array 
(Schuster et al. \cite{sch04}). HERA has nine dual polarization pixels. 
The elements of the array are arranged in a $3\times  3$ matrix, 
with a separation of $24\arcsec$ on the sky between adjacent elements. 
The beam size of the telescope is $\sim$11\arcsec
at these frequencies (Table~4). 
The data were acquired by drifting the telescope in right ascension 
(``on-the-fly'') in the frequency switching mode. The beam 
pattern on the sky was rotated by 18.5 degrees with respect to the right 
ascension axis by means of a K-mirror mounted between the Nasmyth focal plane 
and the cryostat of the heterodyne array. When drifting the telescope in right
 ascension, two adjacent rows are separated by  $8\arcsec$, which results 
in a map slightly under-sampled in declination (the Nyquist sampling step 
is $5\farcs5$). Details about the HERA array and the K-mirror can be found at 
http://www.iram.fr/IRAMES/index.htm. 

We used the WILMA digital autocorrelator, with a spectral resolution of 78 kHz,
as a spectrometer; the resolution was later degraded to obtain a velocity 
resolution of 0.2~km~s$^{-1}$. The observing conditions were typical for the time 
of the year, with system temperatures of 400~K to 500~K. Pointing, which was 
checked every 90 min by scanning across nearby quasars, was found to be 
stable to better than  $2\arcsec$.

Supplementary observations in the \ceio\ \jtwotoone\ and the CS \jtwotoone\, 
\jthreetotwo\ and \jfivetofour\ transitions were carried out towards
the condensations identified in the HERA maps, using the ``standard''  
heterodyne receivers at the IRAM 30-m telescope.
 The weather conditions were very good, with system temperatures of 145 K and 
260 K at 3~mm and 1.3~mm respectively. 

The $^{12}$CO and $^{13}$CO emission extends over scales of several arcmin 
(Fig.~7), comparable to or larger than the first and second error beam 
of the IRAM telescope (see Table~1 in Greve, Kramer, \& Wild, \cite{gre98}). 
Hence the main-beam temperature scale is not a good approximation to the 
intrinsic $^{12}$CO and $^{13}$CO line brightnesses. The antenna temperature scale 
$T_a^{*}$ is a better approximation and we express the CO and $^{13}$CO 
fluxes in this unit. On the 
other hand, the \ceio\ emission is much more compact, and the main-beam 
brightness temperature scale  is a reasonable approximation to the intrinsic 
line brightness. We adopt a value of 0.53 for the main-beam efficiency
at the frequency of the \ceio\ line. 

Table~4 gives a summary of the millimetre observations and the 
efficiencies used.

%
\begin{table}[htp]
\begin{flushleft}
\caption[]{Summary of millimetre line observations. }
\begin{tabular}{|l|c|c|c|c|} \hline
Line & Frequency & Beam width & $B_{\rm eff}$ & $F_{\rm eff}$ \\  
     &  (GHz)    & (arcsec)  &               &         \\ \hline
CS $\jtwotoone$    & 97.98097  & 24 & 0.77  & 0.95 \\
CS $\jthreetotwo$  & 146.96905 & 16 & 0.69  & 0.93 \\
$\ceio~\jtwotoone$ & 219.56038 & 11 & 0.53  & 0.90 \\
$\thco~\jtwotoone$ & 220.39872 & 11 & 0.52 & 0.90\\ 
$\twco~\jtwotoone$    & 230.53800 & 11 & 0.52 & 0.90 \\
CS $\jfivetofour$  & 244.93563 & 10 & 0.54 & 0.91 \\
\hline 
\end{tabular}
\end{flushleft}

\end{table}

%
\begin{figure*}[htp]
\includegraphics[width=180mm]{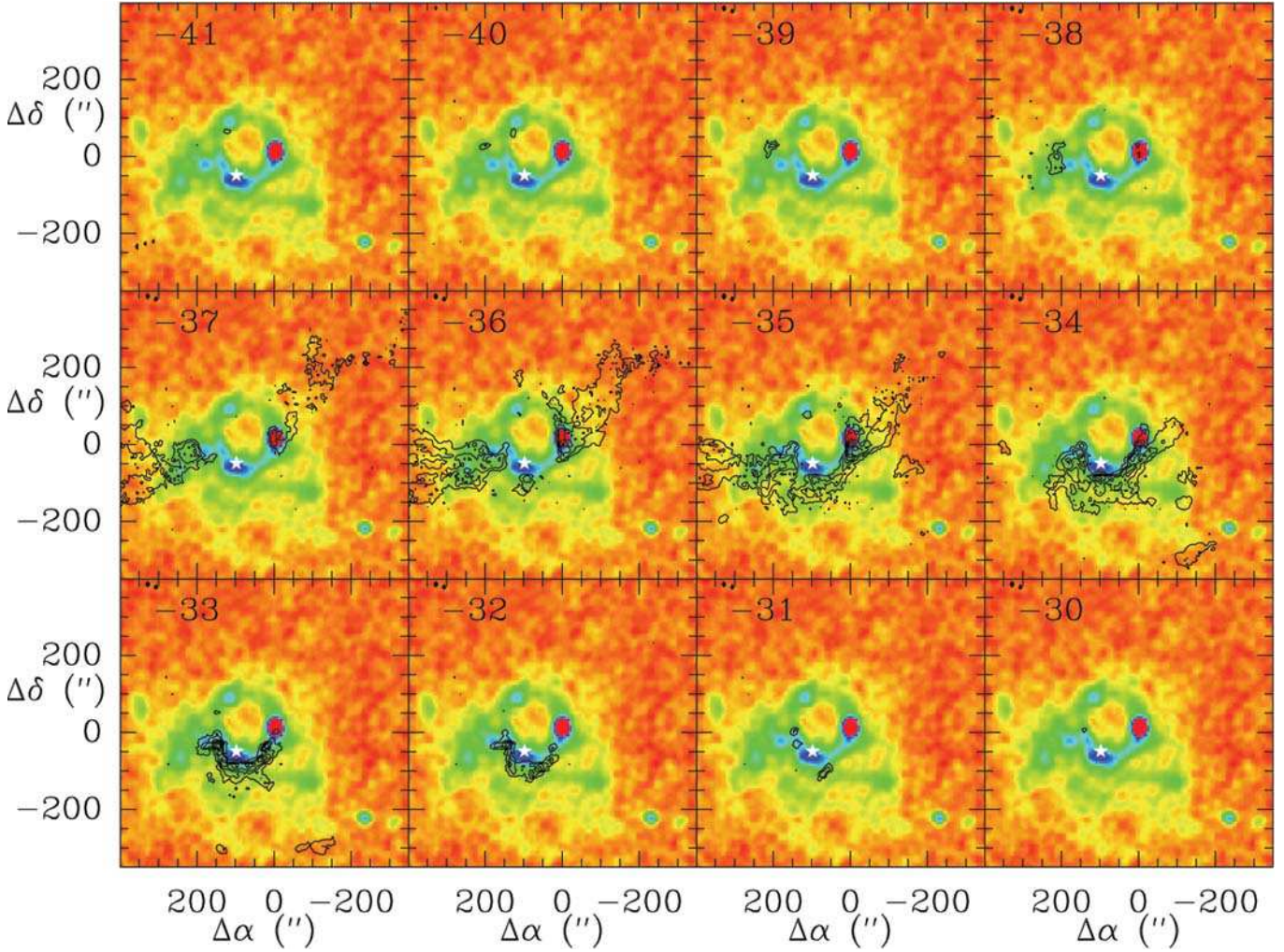}
\caption{Channel maps of the \twco\ \jtwotoone\ emission as observed with HERA 
and integrated over velocity intervals of $1\kms$; the central velocity of each bin 
is marked in the upper left corner of each panel. The first contour 
is at 1~K~km~s$^{-1}$ and the contour interval is 2~K~km~s$^{-1}$. The coloured background 
is the MSX emission at 8.3~$\mu$m. The 0,0 position is that of the 
MSX point source (Table~1).}
\end{figure*}

\subsection{Distribution and kinematics}

In all the maps presented hereafter, the coordinates are expressed in 
arcsecond offsets with respect to the MSX point source.
The distribution of the integrated \twco\ emission as a function 
of velocity is shown in Fig.~7. The CO line traces a diffuse 
filamentary cloud which extends southeast--northwest. The velocity of this 
filament is between $-36$ and $-37\kms$, hence a few kilometres per 
second more positive 
than that of the ionized gas. A bright and thin half-ring structure of molecular gas 
is very clearly associated with the photo-dissociation region. This 
half-ring follows the ring of MSX emission at 8.3~$\mu$m which surrounds 
the brightest part of the ionized region. It lies at the 
back of the \Hp\ region, as no corresponding extinction of the optical 
nebular emission is observed in its direction. This is consistent with the 
observed velocity field, which shows that the 
molecular gas ring is redshifted with respect to the molecular filament 
and the ionized gas.  

The ring is fragmented into at least five condensations (Fig.~8). From west to east 
we have condensation 1, observed in the direction of the MSX point source and the 
UC \Hp\ region (see Sect.~5), with a mean velocity of $-$35.5~km~s$^{-1}$; 
condensations 2 and 3 with a velocity of $-$33.5~km~s$^{-1}$; condensation 4 
with a velocity of $-$32.5~km~s$^{-1}$. Further to the east lies condensation
5 with a velocity of $-$37.5~km~s$^{-1}$.  

The velocity varies along the half-ring structure: condensations 1 and 5, situated 
on opposite ends of this structure, have velocities not very different 
from that of the filament; the half-ring, at the rear of the 
Sh2-212 \Hp\ region, is expanding  with a velocity of a few kilometres per second  
with respect to the filament. The ionized gas flows away from 
the filament in the opposite direction (see Fig.~13).
 
%
\begin{figure*}[htp]
\includegraphics[width=180mm]{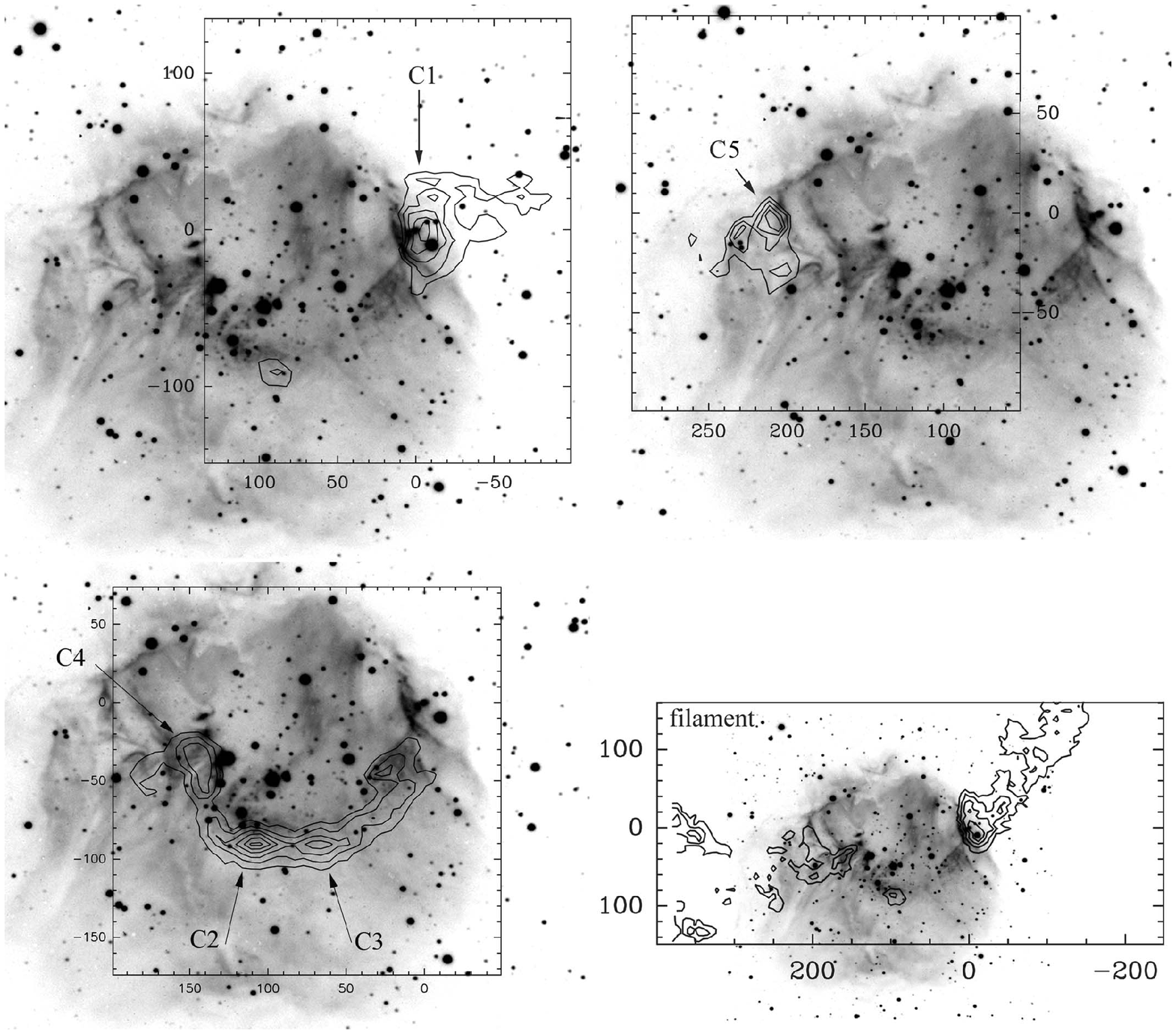}
\caption{{\bf Top Left:} Condensation C1, showing the $^{13}$CO(2-1) emission integrated 
between $-36.1\kms$ and $-35.1\kms$. The first contour and 
the step correspond respectively to 20\% and 15\% of the peak brightness 
($9.6\K\kms$). {\bf Bottom Left:} Condensations C2, C3, and C4, showing 
the $^{13}$CO(2-1) emission 
integrated between $-34.0\kms$ and $-32.7\kms$. The first contour and 
the step are respectively  20\% and 15\% of the peak brightness ($15.7\K\kms$).  
{\bf Top Right:} Condensation C5, showing the $^{13}$CO(2-1) emission 
integrated between 
$-38.4\kms$ and $-36.6\kms$. The first contour and the step are respectively 
35\% and 15\% of the peak brightness ($3.8\K\kms$). {\bf Bottom Right:} Filament, 
with the {$^{13}$CO(\mbox 2-1)} emission integrated between $-36.8\kms$ and $-35.9\kms$. 
The peak is at 6.6~K~km~s$^{-1}$. The first contour and the step are 
respectively 20\% and 20\% of the peak brightness.}
\end{figure*}

Fig.~8 compares the \thco\ emission, especially the locations of the 
condensations, with the \SII\ emission of the ionized gas. Fig.~8 shows that 
the bright rim present at the northwest border of Sh2-212, which harbours the 
UC \Hp\ region and star no.~228, is the ionized border of condensation 1. 
Thus condensation 1 appears as the remains of the parental core in 
which the massive YSO no.~228 formed and subsequently ionized a UC \Hp\ 
region. Furthermore, several  molecular substructures have counterparts in 
the \SII\ image, as substructures of the ionization front. For example, 
condensation 5 is situated behind an IF traced by \SII\ emission.

%
\begin{figure}[htp]
\includegraphics[width=90mm]{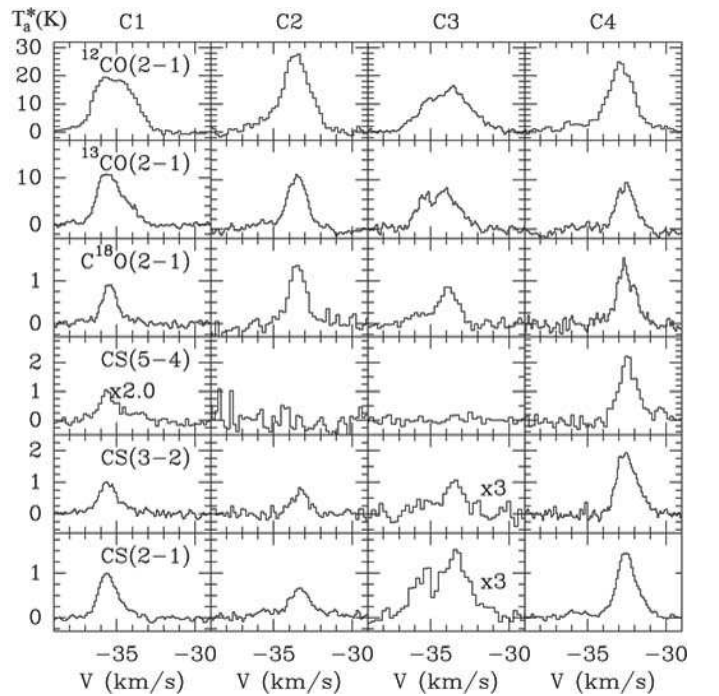}
\caption{Molecular lines observed towards the molecular fragments C1 to C4.}
\end{figure}

\subsection{Physical conditions}

%
\begin{table*}
\begin{flushleft}
\caption[]{Properties of the molecular gas condensations}
\begin{tabular}{|c|c|c|c|c|c|c|c|c|}\hline
Condensation  & Offset Position & $\Delta v^1$ & 
Core Dimensions & $N(\htwo)^{\rm peak,1}$ & $n(\htwo)^1$ & $N$(CS) & $n(\htwo)^{\rm peak,2}$ & 
Fragment Mass   \\
      &   (\arcsec) & ($\kms$)  & (\arcsec) & ($\cmmd$) & ($\cmmt$) & ($\cmmd$) & ($\cmmt$) & ($\msol$) \\ 
\hline 
1 & ($-6,-5$)  & 1.1 & $25 \times 8$  &  
4.3(21) & 1.0(4) & 6.5(12) & 4(5)  & 220 \\
2 &($+108,-90$) & 1.2 & $25 \times 12$ & 
6.7(21) & 1.0(4) & 3.8(12)  & 4(5) & 63 \\
3 & ($+68,-90$) & 1.2 & $25 \times 10$ & 
4.3(21) & 8.2(3) & 2.7(12) & 1.8(5) & 45 \\ 
4 & ($+136,-30$) &1.0& $15 \times 10$ &
7.2(21) & 2.5(4) & 1.2(13)  & 1.0(6) & 61 \\
5 & ($+210,-5$)& $2.0^3$& $16 \times 16$ & 
2.7(21) & \_ & \_  & \_ & 38 \\
\hline
\end{tabular}
\end{flushleft}
$^1$~determined from \ceio\ data. \\
$^2$~determined from CS data.\\
$^3$~from \thco\ data. \\
\end{table*}

\subsubsection{The kinetic temperature}

$^{12}$CO brightness temperatures of 20--30~K are observed 
along the molecular gas ring and the cores. 
Towards C1, the maximum of \twco\ brightness is observed at the offset position 
$(-3\arcsec,-1\arcsec)$, where $T_{\rm a}^{*}= 24\K$. This implies a kinetic
temperature $T_{\rm k}\approx 27\K$. 
The peak brightness temperatures in the \thco\ and \ceio\ \jtwotoone\ 
transitions are $11\K$ and $1.5\K$, compatible with 
opacities $\tau_{13}= 0.58$ and $\tau_{18}= 0.08$, respectively, adopting 
standard relative abundances $\rm [^{12}CO]/[^{13}CO]= 60$ and 
$\rm [^{13}CO]/[C^{18}O]= 8$.  Hence  the $\thco$ \jtwotoone\ transition is 
moderately optically thin. The \ceio\ line traces a gas column density 
$N(\htwo)=4.3\times 10^{21}\cmmd$ at the intensity peak of C1. 
Calculations of the \ceio\ 
line excitation in the large velocity gradient (LVG) approximation show that the 
opacity of the line is consistent with a gas kinetic temperature of $30\K$, 
as estimated above. A similar conclusion was derived for C3. Hence we adopt
a kinetic temperature of $30\K$ for the condensations,  in what follows.

Inside the filamentary cloud, away from the \Hp\ region, the \twco\ and \thco\  
\jtwotoone\ ~brightness temperatures are much lower, typically a few kelvins  
for both lines. At the offset 
position $(-120\arcsec,+80\arcsec)$, we measure $T_B(\twco)= 8.2\K$ 
and $T_B(\thco)= 2.1\K$, which implies an opacity $\tau_{13}= 0.32$, adopting 
the same standard abundance ratio.  
Both line intensities and opacities are accounted for by a gas layer of 
column density $N(\htwo)= 9\times 10^{20}\cmmd$ at about $14\K$.

\subsubsection{Masses and Densities}

The masses of the molecular gas fragments were derived from the \thco\ data, 
in the optically thin limit. For condensation C1 the contour at half power 
of the \thco\ \jtwotoone\ emission delineates a condensation of 
$56 \arcsec \times 28\arcsec$, oriented north-south, centred at offset 
position $(-6\arcsec,-1\arcsec)$. The total mass of the condensation is 
obtained by integrating over the 20\% peak contour, from which we determine 
$M= 220\msol$. 
A similar procedure was applied to the five condensations. The results 
are summarised in Table~5.

The half shell of collected molecular material has a mass  
$\leq$720\,\msol, as estimated by integrating the whole 
$\thco\ \jtwotoone$\  emission in the ring. 

The mean density in a fragment, $n(\htwo)$, is derived as 
$N(\htwo)/\sqrt(ab)$ where $a$ and $b$ are the major and minor axes of 
the condensation. We have taken into account the dilution in the 
main beam, by applying a correction factor $(ab)/(ab+\theta_{\rm beam}^2)$, 
where $\theta_{\rm beam}$ is the beam width. These mean densities are given 
in Table~5.

Emission of the high-density gas was traced by the millimetre lines of CS (Fig.~9).
The emission of the lower transition \jtwotoone\ is detected along the 
filament, whereas the \jthreetotwo\ and \jfivetofour\ emissions are more 
compact. The \jfivetofour\ transition was detected only in C1 and C4, arising from 
a small unresolved region. The \jtwotoone\ and \jthreetotwo\ lines intensities 
are typically 1 to $2\K$ at the brightness peak of the condensations. The 
lines are typically $1\kms$ wide. 

Estimates of the \htwo\ density in the fragments were obtained by modelling 
the millimetre CS line emission, in the LVG approximation.
Analysis of the \jtwotoone\ and the \jthreetotwo\ transitions at the 
brightness peak 
indicate typical densities $n(\htwo)\simeq$ 2--4 $\times 10^5\cmmt$ in the cores, 
and up to $1.0\times 10^6\cmmt$ towards C4. Note that these densities 
are much higher than the mean densities estimated from the \ceio\ emission; 
thus the CS material has a small filling factor.

\section{Radio observations}
We observed Sh2-212, in the direction of the MSX point source, with
the Very Large Array (VLA) on 2005 November 16.  We searched for
emission from methanol at 44~GHz (7~mm), water at 22~GHz (1.3~cm),
ammonia in the (1,1) and (2,2) lines at 23~GHz, and neutral hydrogen
at 21~cm.  The array was in its most compact configuration, providing
spatial resolutions of $1\farcs5$, $3\farcs3$, and $45\arcsec$ in the 7~mm,
1.3~cm, and 21~cm bands, respectively. Each molecular transition 
was observed with a distinct combination of bandwidth and number of 
channels, resulting in spectral resolutions of 0.3~km~s$^{-1}$ for 
methanol and water, and 0.6~km~s$^{-1}$ for ammonia. For all three 
molecules the total bandwidth corresponds to about 40~km~s$^{-1}$, 
and is centred at $-35$~km~s$^{-1}$. Further observational 
details will be provided in a forthcoming paper.
We did not detect any of the molecular transitions to the 
5~$\sigma$ levels
of 90, 33, and 10 mJy~beam$^{-1}$ for the methanol, water, and
ammonia lines, respectively.  The neutral hydrogen line at 21~cm was
detected, and is discussed below.\\

The absence of 44~GHz methanol masers is not necessarily surprising.
Methanol masers are known to be associated with high-mass star 
formation (e.g., Ellingsen \cite{ell06}); however, type II methanol masers 
(such as those producing the 6.7 and 12.2~GHz lines) are thought to be pumped by the 
radiation field of YSOs, and hence are closely linked with the star 
formation process.  Type I masers (such as those producing the 44~GHz line) are 
thought to be collisionally pumped, and may not be directly associated 
with massive YSOs.  More surprising is the absence of water masers, 
which are nearly ubiquitous in star-forming regions. Water masers are 
known to be variable, and a possible explanation for our  
non-detection is that such masers are present but currently quiescent.  
Other explanations for the non-detection may have 
implications for the star formation process. For example, water 
masers are typically thought to arise in outflow and/or accretion 
processes, which may be absent in YSO no.~228.

Our 5~$\sigma$ 
detection level of 10 mJy beam$^{-1}$ for the (1,1) and (2,2) ammonia 
lines allows us to estimate an upper limit for the gas column density. 
Assuming optically thin emission, with a 30~K excitation temperature, 
and a beam filling factor of one, our observations could detect an 
ammonia column density of about $10^{14}$~cm$^{-2}$.  Adopting an 
ammonia abundance of $10^{-8}$ indicates a detection limit of H$_2$ 
column density of about $10^{22}$~cm$^{-2}$.  This is marginally 
higher than the several times $10^{21}$~cm$^{-2}$ column densities 
reported in Table 5; thus our non-detection of ammonia is consistent 
with the column densities inferred from the CO observations. 
Hot molecular cores, a common tracer of young, high-mass star 
formation, with column densities of $10^{23}$--$10^{24}$~cm$^{-2}$, 
are clearly absent. \\

Centimetre continuum data of Sh2-212 were obtained from the VLA data
archive of programmes AF346 and AR390, at 1.46 GHz and 8.69~GHz 
respectively. These data were calibrated and imaged using standard
procedures for continuum data.  The AF346 observations were made in
1998 in the C and B configurations.  The lower resolution
(16\arcsec$\times$13\arcsec) C-array data imaged both the Sh2-212
region and a compact source to the northwest.  The resolution was  
too low for a reliable determination of the compact source's
parameters; the higher resolution (5\arcsec$\times$4\arcsec) B-array
data were used for this purpose.  The AR390 observations were made in
1997 in the D configuration, and provided an angular resolution of
10\arcsec$\times$7\arcsec.  Sh2-212 was too large to be imaged by
these data, but the compact source parameters were reliably determined 
to be $11\pm1$~mJy at 1.46~GHz and 
$7.6\pm1$~mJy at 8.69~GHz. Gaussian fits to the source size and
position yield deconvolved major and minor axes of 3\arcsec, and a J2000 
position 04$^{\rm h}$40$^{\rm m}$27$^{\rm s}$.2, +50\degr28\arcmin29\arcsec, 
i.e. coincident with the MSX source G155.3319+02.5989 (see Table 1).
Although no ammonia was detected in our 2005 observations, we used the
central 75\% of the 3.1~MHz bandwidth to form a 1.3~cm continuum
image, presented in Fig.~10. The compact source was detected in 
this image, with a flux density of $8.9\pm1.2$~mJy.

For the UC \Hp\ region, the flux densities at 1.46, 8.69, and 23.7~GHz show a 
flat spectrum, indicative of an optically thin \Hp\ region.  All three 
observations indicate an ionizing photon flux of log $N_{\rm Lyc}>46.5$, or the
 equivalent of a B1 or earlier star (Smith et al.~\cite{smi02}). The 3\arcsec\ size 
corresponds to 0.095~pc at a distance of 6.5~kpc.  Assuming a spherical \Hp\ 
region of this diameter, the flux densities indicate an rms electron
density of $3.2 \times 10^3$~cm$^{-3}$, with a total mass of ionized gas
of 0.05~$M_\odot$.\\

%
\begin{figure}[htp]
\includegraphics[width=90mm]{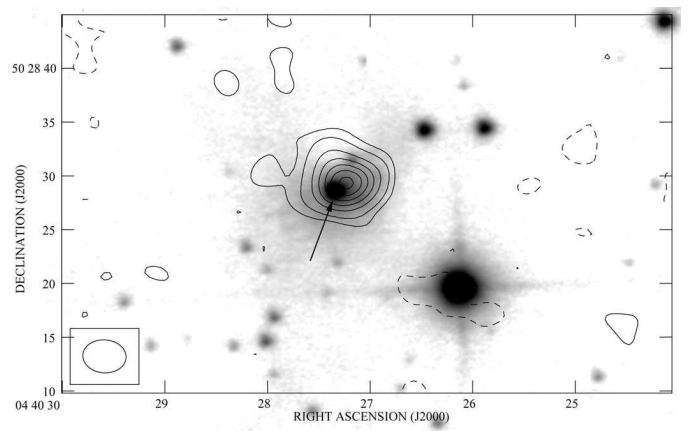}
\caption{ Radio continuum emission (contours) 
superimposed on the $H$ image (grey scale).  Star no. 228 is indicated by an arrow. 
The contours show 1.3 cm continuum emission from the UC \Hp\ region 
coinciding in direction with the
MSX point source and the star.  The angular resolution of the radio image
is $3 \farcs 3$ (indicated in the lower left corner) and the contour
levels are $-$15, 15, 20, 25, 30, 40, 45 mJy~beam$^{-1}$.}
\end{figure}

Sh2-212 was also observed with the VLA in the 21~cm line of
neutral hydrogen.  The observations were made with a 1.56~MHz
(330 km~s$^{-1}$) bandwidth and 255 channels of 6.1~kHz (1.3 km~s$^{-1}$)
each.   Data reduction followed standard VLA spectral line procedures.
After the external flux and phase calibration, continuum emission was
subtracted from the $uv$ data and an image cube was formed. Very extended
\HI\ emission was present in the field while the \HI\ emission on size scales
similar to those of Sh2-212 was relatively weak.   To optimise the imaging
toward the \HI\ associated with Sh2-212, the data were re-imaged, removing
the shortest 0.3~k$\lambda$ baselines (to suppress the extended emission)
and averaging adjacent channels (to improve the signal-to-noise ratio).
The resulting image cube has an angular resolution of $50'' \times 41''$ and
a spectral resolution of 2.6~km~s$^{-1}$.  \HI\ emission was found in 
three adjacent channels, from
$-40$~km~s$^{-1}$ to $-48$~km~s$^{-1}$. Three distinct \HI\ condensations
were found, all lying (in projection) at the edge of the \Hp\ region.  
A contour plot of this emission is shown in Fig.~11. 
Using the peak brightness temperature (in a 45$''$ beam) we can 
calculate the column density over the central 1.4 pc of each clump. 
Assuming optically thin emission, so that the observed line temperature 
is approximately equal to the spin temperature times the optical depth, 
we calculate the column density as 
$N_{\rm HI}$(cm$^{-2})=1.82 \times 10^{18}~(T_L$/K)($\Delta V$/km~s$^{-1}$). 
All three clumps have peak line temperatures of 20--25~K, and linewidths of 
4--5~km~s$^{-1}$. Hence, for all three we find 
column densities of about 1.5$\times 10^{20}$~cm$^{-2}$ (within a factor of two). 
Assuming a spherical 
geometry for the central region of each clump implies hydrogen densities 
of about 35~cm$^{-3}$.
%
\begin{figure}[htp]
\includegraphics[width=90mm]{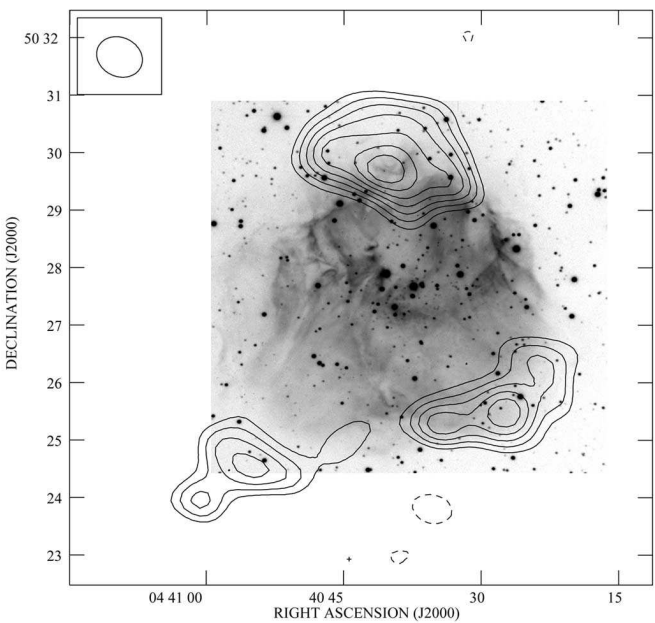}
\caption{\HI\ emission associated with Sh2-212, integrated between
$-40$~km~s$^{-1}$ and $-48$~km~s$^{-1}$. The radio angular
resolution ($50'' \times 41''$) is indicated in the upper left corner.
Contour levels are $-15$, 15, 20, 25, ... 50 mJy~beam$^{-1}$~km~s$^{-1}$.
The \HI\ contours are superimposed on the \SII\ image.}
\end{figure}~

\section{Discussion}
\subsection{A massive young stellar object exciting a UC H\,{\scriptsize\it II} region}

Star no.~228, associated with a reflection nebulosity, is a 
massive young stellar object:

$\bullet$ It is a luminous near-IR source with a near-IR excess indicative of 
the presence of nearby dust, probably associated with a disk 
(Lada \& Adams~\cite{lad92}).

$\bullet$ Its spectral energy distribution (SED) rises strongly in the IR. 
Table~6 gives the flux measurements between 1.25~$\mu$m and 21.3~$\mu$m. We have used 
the Web-based SED fitting tool of Robitaille et al.~(\cite{rob07}; 
http://caravan.astro.wisc.edu/protostars/ ) to interprete that of YSO no.~228 
and its associated nebulosity.  Several parameters 
of the model are not well constrained, but some are. A 
strong conclusion is that the central stellar object is hot 
($T_*=30000 \pm 1000$~K) and  massive ($M_*\sim14$\msol), and that the YSO is luminous 
(total luminosity $14000 - 20000$\lsol); the disk and the envelope have 
similar luminosities (more uncertain). Also, the YSO is seen edge-on (inclination 
angle $\sim$87\degr). Fig.~12 shows the SED of YSO no.~228 and the five best-fitting 
models.

Fig.~10 shows the radio-continuum emission of the UC \Hp\ region at 1.3~cm , as 
contours  superimposed on an $H$ image. Star no.~228 lies at the centre 
of the UC \Hp\ region (see also Table~1).  Thus star no.~228 is very 
probably the exciting star of the UC \Hp\ region.

The SED of object no.~228 indicates that it is probably in an evolutionary stage 
between Class~I and Class~II, with both an envelope and a disk. This object 
is consistent with the evolutionary models of massive stars, formed by accretion, 
as described by Beech \& Mitalas (\cite{bee94}) and by Bernasconi \& Maeder 
(\cite{ber96}). This object of $\sim$14\,\msol\ has reached the main sequence, 
and hence is burning hydrogen in its centre. But it is still accreting material 
and increasing its mass. Presently its accretion rate is not high enough 
to prevent the formation of an ionized region (Walmsley, \cite{wal95}).

This massive YSO, which is observed at its place of birth inside the 
parental condensation, does not 
seem to belong to a populous cluster. Only three low-brightness stars are 
observed nearby, at less than 0.11~pc (with $K$ magnitudes $\geq16.1$).
{\it It is therefore a good candidate for being a massive star born either in isolation 
or in a very small cluster}. As such, it deserves further high-resolution,  
high-sensitivity imaging and spectroscopic observations to ascertain its nature 
and to detect any nearby deeply-embedded objects.

\begin{table}[htp]
\caption{Spectral energy distribution of star no.~228, a massive YSO}
\begin{tabular}{ccl}
 \hline\hline
 Wavelength & $F_{\nu}$ & origin \\
 ($\mu$m)  & (Jy)  & \\
 \hline
 1.215 & 0.629 10$^{-2}$ & $J$ star alone \\
 1.215 & 1.268 10$^{-2}$ & $J$ star + nebulosity \\
 1.65 & 1.049  10$^{-2}$ & $H$ star alone \\
 1.65 & 2.006  10$^{-2}$ & $H$ star + nebulosity \\
 2.18 & 1.268  10$^{-2}$ & $K$ star alone \\
 2.18 & 3.112  10$^{-2}$ & $K$ star + nebulosity \\
 8.28 & 2.27  & MSX star + nebulosity \\
 12.13 & 2.91 & MSX star + nebulosity \\
 14.65 & 4.94 & MSX star + nebulosity \\
 21.3 & 35.34 & MSX star + nebulosity \\
 \hline \\
\end{tabular}
\end{table}
%
\begin{figure}[htp]
\includegraphics[width=90mm]{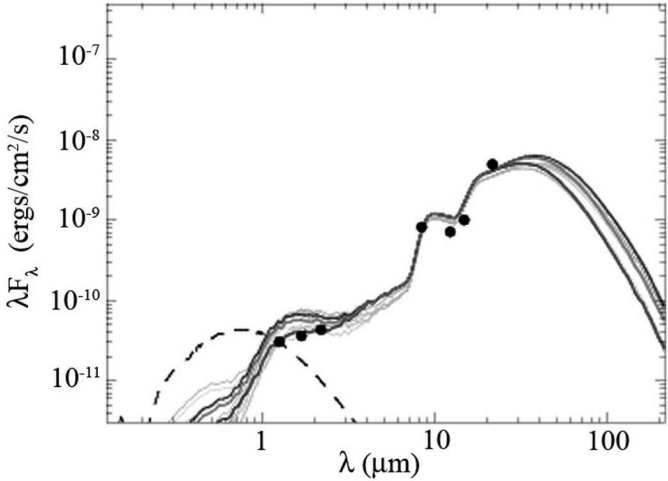}
\caption{Spectral energy distribution of star no.~228, a massive YSO. 
Filled circles are the fluxes listed in Table~6. The five best 
fitting models obtained using the web-based tool of 
Robitaille et al.~(\cite{rob07}) are presented.}
\end{figure}

\subsection{The age of the UC H\,{\scriptsize\it II} region}

The UC \Hp\ region formed and evolved in condensation 1. Let us assume that 
it formed in a 
uniform medium of density $10^5$~cm$^{-3}$. The exciting star, emitting 
10$^{46.5}$ ionizing photons per second, very quickly formed an 
ionized region of radius 
0.0046~pc, which later expanded. According to Dyson \& 
Williams~(\cite{dys97}, sect.~7.1.8) the present radius of the UC \Hp\ region, 
0.0475~pc, corresponds to an age of 13\,500~yr, and a  
density in the ionized gas of 3050~cm$^{-3}$, which is very close to 
that which is observed. Also, the pressure equilibrium with the surrounding 
medium has  not yet been reached (the expansion velocity 
being $\sim$2~km~s$^{-1}$).

Thus this UC \Hp\ region is very young -- much younger than Sh2-212 (see 
Sect.~6.4). Our non-detection of water maser emission is a little surprising, 
and higher-sensitivity observations are needed.

Note, however, that the age of the \Hp\ region is not that of the massive 
YSO no.~228, which is much older, having evolved during a long time 
before being able to ionize the surrounding gas.

\subsection{The morphology of the Sh2-212 complex}

The present distribution of the molecular material 
indicates that Sh2-212 probably formed in a filament. Was this medium  
turbulent? 
One morphological aspect of this region, the perfectly  
spherical shape of the ionized region, at both optical and radio wavelengths, 
seems to indicate that the level of turbulence, if any, is low. 
We suggest (see Fig.~13) that the exciting star of Sh2-212 formed inside the 
molecular filament 
and that the \Hp\ region first expanded inside this filament. When 
the ionization front reached the border of the filament, the \Hp\ region 
opened on the outside (a low-density inter-filament medium), 
thus creating a champagne flow. This explains both the 
velocity field (the ionized gas flowing away from the molecular cloud, more 
or less in the direction of the observer, at a few kilometres per second), 
and the shape of the \Hp\ region 
and its photodissociation region containing the PAHs (a bright ionized region  
surrounded by the PAH emission ring observed by MSX, and a more diffuse extension 
of the ionized gas).

%
\begin{figure}[htp]
\includegraphics[width=90mm]{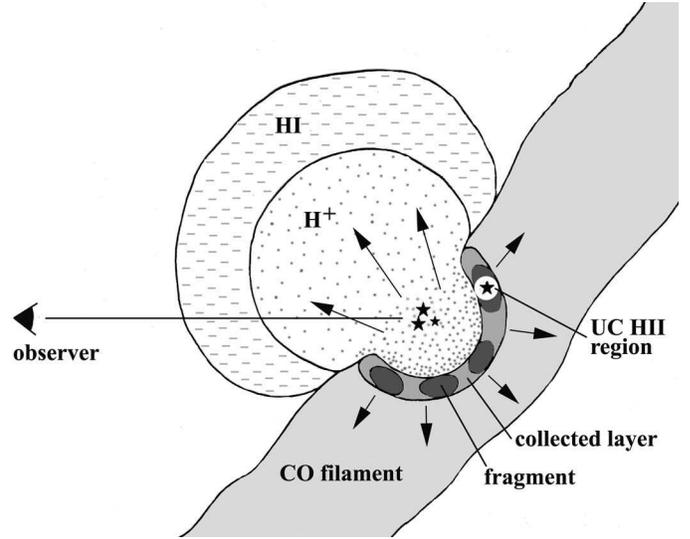}
\caption{Morphology of the Sh2-212 complex.}
\end{figure}

During its supersonic expansion inside the molecular filament, the 
ionization front was 
preceded by a shock front on the neutral side, and neutral material 
accumulated between the two fronts. This collected material forms the 
thin molecular half-ring which surrounds the 
brightest part of the \Hp\ region. This collected layer forms only half 
a shell, adjacent to the molecular filament, and is in expansion with a velocity 
of 4--5 km~s$^{-1}$ with respect to the molecular filament. On the 
other side the \Hp\ region 
is surrounded by low-density atomic material; this is the origin 
of the \HI\ emission observed at the periphery of \mbox{Sh2-212} (Fig.~11). This atomic 
material is receding from the molecular filament with a velocity similar 
to that of the ionized gas. The low density of this material is confirmed 
by the very low, if any, local extinction in front of the ionized gas. 
Such atomic hydrogen rings have been observed in other complexes, 
for example around the spherical  
\Hp\ region Sh2-219 (Roger \& Leahy~\cite{rog93}, Deharveng 
et al.~\cite{deh06}, Fig.~4), and around Sh2-217 (Roger \& Leahy~\cite{rog93}, 
Brand et al.~in preparation). 

Dale et al.~(\cite{dal05}) have  simulated the photoionizing feedback 
of a massive  star on a turbulent molecular cloud. Fig.~14 shows the striking 
similarity between the morphologies of the observed Sh2-212 molecular cloud and of 
the simulated cloud. The right panel -- simulation -- shows the column density of the 
neutral gas; the ionized region lies in the central hole. The left panel -- observations -- 
shows the $^{12}$CO emission integrated over velocity (the intensity is 
proportional to the column density, except for regions optically thick 
along the line of sight); here again the ionized region lies 
inside the central hole. The simulation concerns a turbulent cloud of relatively 
low density; a more chaotic morphology is obtained in the case of a denser 
turbulent cloud (Dale et al.). 

%
\begin{figure*}[htp]
\includegraphics[width=180mm]{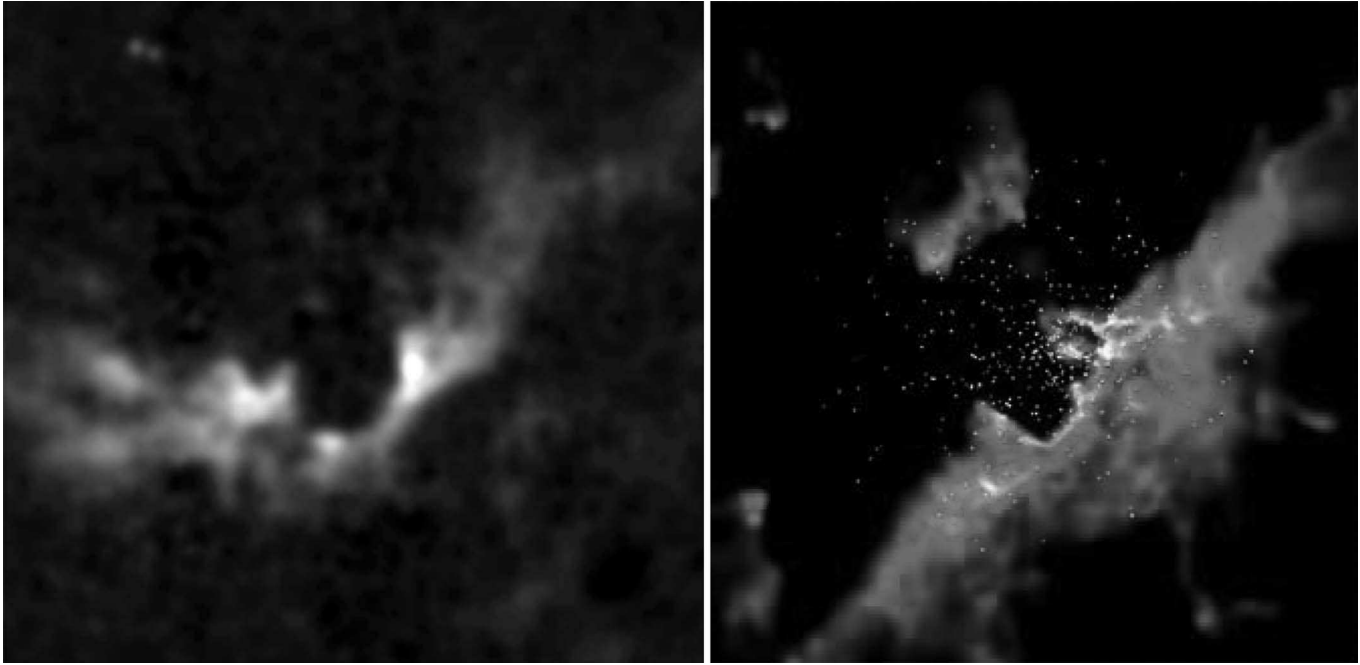}
\caption{Confrontation of observations and simulations. {\bf Left:} $^{12}$CO 
emission, integrated over the velocity. {\bf Right:} simulation of a turbulent 
cloud illuminated by the UV radiation of a massive star 
(Dale et al.~\cite{dal05}, their Fig.~16).}
\end{figure*}

\subsection{Star formation history}

It is almost impossible to obtain direct evidence of sequential star 
formation. The age of the evolved Sh2-212 is very uncertain 
due to our lack of knowledge about 
the density structure of the original medium in which this \Hp\ region 
formed and evolved. Only indirect evidence is available.

It is difficult to explain the origin of the thin circular half-ring 
of molecular material which surrounds the brightest part of Sh2-212 other than by 
material collected during the expansion of this \Hp\ region. It is presently 
fragmented; at least five condensations are present along the ring. This is a 
good illustration of the collect and collapse process. The most massive 
condensation contains a massive young stellar object exciting a UC \Hp\ region. 
This UC \Hp\ region is very young -- much younger than Sh2-212. Thus 
massive star-formation has been triggered by the expansion of the Sh2-212 \Hp\ region, 
via the collect and collapse process. Here again this process seems able to 
form  massive objects, as observed on the borders 
of Sh2-104 (Deharveng et al.~\cite{deh03}) and RCW~79 (Zavagno et al.~\cite{zav06}). 
Also, a few lower mass YSOs have formed in the collected layer, for example 
in the direction of C2.

In order to compare with the predictions of the collect and collapse model 
of Whitworth et al.~(\cite{whi94}), we need to know three parameters: 1) the Lyman 
continuum photon flux: we adopt 10$^{49}$ ionizing photons per second, 
corresponding to the O5.5V--\,O6V exciting star (Martins et al.~\cite{mar05}); 
2) the velocity dispersion in the collected layer: we measure a 
FWHM $\sim$1~km~s$^{-1}$ at the condensation peaks (Table~5), 
corresponding to a velocity dispersion $\sim$0.4~km~s$^{-1}$. The condensations 
are possibly collapsing, and thus a lower value, in the range 
0.2--0.3~km~s$^{-1}$, seems reasonable for the collected layer before collapse; 
3) the density of the neutral material into which the \Hp\ region 
evolved: a density $\sim$500~atom~cm$^{-3}$ allows formation of a collected layer 
(half a shell, of internal radius 2~pc) with a mass of about 750~$M_{\odot}$, 
as observed. With these figures the model predicts the fragmentation 
of the collected layer after 2.2--2.8~Myr . It predicts the formation of 
fragments with a mass in the range 30--140~$M_{\odot}$, in agreement 
with the observations, separated by some 1.1--2.2~pc, again in 
agreement with the observations. But the radius of the \Hp\ region at the time 
of the fragmentation should be in the range 8.5--10.0~pc, much larger than 
the present radius of 2~pc. The fact that the \Hp\ region shows a 
champagne flow (and thus did not evolve in a homogeneous medium) probably 
explains why this model does not account for the observations. The possible 
presence of a magnetic field is an additional difficulty because, as demonstrated by 
Krumholz et al.~(\cite{kru06}), it results in a non-spherical expansion.

The star formation efficiency can be estimated. 
Condensation 1 has a mass $\sim$220\,\msol\ and has formed a 
(possibly isolated) B1V star; assuming for this star a mass of 14~\msol, we obtain 
a star formation efficiency of 5\%.

\section{Conclusions}

The optical Galactic \Hp\ region Sh2-212 appears in the visible as a 
spherical \Hp\ region  
around its O5.5V exciting star. The near-IR observations show that a 
rich cluster lies at its centre. A bright stellar object, no.~228, 
presenting a near-IR excess, and associated with a reflection nebulosity, 
lies on the border of Sh2-212, behind a bright rim. The MSX image at 
8.3~$\mu$m shows a bright point source in the direction of object no.~228,   
and radio continuum observations show the presence of a UC \Hp\ region in this exact 
direction. Sh2-212 lies in the middle of a molecular filament. Millimetre 
observations show that a thin molecular half-ring 
structure surrounds the brightest part of Sh2-212 at its back, and is expanding. 
This molecular layer is fragmented. The most massive fragment ($\sim$200\,\msol) is 
associated with object no.~228. The SED of object no.~228 shows that it 
is a massive YSO of about 14 \msol , hence able to ionize 
the UC \Hp\ region. Low-density atomic hydrogen is detected at 
the periphery of the low-density ionized region.\\

We have tried to understand the star formation history in this region. 
Sh2-212 first formed and evolved inside 
a molecular filament. During its expansion neutral material was collected 
between the IF and the SF, as predicted by the collect and collapse process. 
This layer was then fragmented, and a second-generation massive star 
(no.~228) formed inside a massive fragment, ultimately ionizing a   
second-generation \Hp\ region. In a next stage, the IF bounding the 
Sh2-212 \Hp\ region reached the limits of the molecular filament, and the 
ionized region opened towards the low-density inter-condensation gas, 
creating a champagne flow. Presently, the Sh2-212 \Hp\ region is 
surrounded on one side by the dense collected molecular material, 
and on the other side by low-density atomic material.\\

The Sh2-212 \Hp\ region is, after Sh2-104 and RCW~79, one more example of 
massive-star formation by the collect and collapse process. It is a  
very special region for the following reasons:

- The massive YSO, no.~228, seems to have formed in isolation, or 
in a very small group. If this is confirmed, no.~228 is a very uncommon object.

- The layer of collected material is very well defined 
(circular, bright, thin, expanding); the fact that it 
survives in an inhomogeneous medium is especially interesting. 
This demonstrates that the collect and collapse process can work in 
a non-homogeneous medium, possibly in a turbulent one. 

\begin{acknowledgements}

We gratefully thank D.~Gravallon and S.~Ilovaisky for the \Ha\ and \SII\ frames they 
obtained for us at the 120-cm telescope of the Observatoire de Haute-Provence, 
and M.~Walmsley for constructive comments and questions. 
This work has made use of Aladin and of the Simbad astronomical database operated 
at CDS, Strasbourg, France. We have used data products from the Midcourse 
Space EXperiment and from the Two Micron All Sky Survey, obtained through the 
NASA/IPAC Infrared Science Archives. S. K. thanks the Laboratoire d'Astrophysique 
de Marseille and the Universit\'e de Provence for hosting him while some of 
this work was done.

\end{acknowledgements}


{}

\end{document}